\documentclass{article}
\usepackage{ijcai25}

\usepackage{times}
\usepackage{soul}
\usepackage{url}
\usepackage[hidelinks]{hyperref}
\usepackage[utf8]{inputenc}
\usepackage[small]{caption}
\usepackage{graphicx}
\usepackage{amsmath}
\usepackage{amsthm}
\usepackage{booktabs}
\usepackage{algorithm}
\usepackage[noend]{algorithmic}
\usepackage[switch]{lineno}
\usepackage{pgfplots}

\title{Dynamic Coalitions in Games on Graphs with Preferences over Temporal Goals}

\author{
	And Kaan Ata Yilmaz$^1$
	\and
	Abhishek Kulkarni$^1$ \and
	Ufuk Topcu$^1$ \\
	\affiliations
	$^1$ The University of Texas at Austin
	\emails
	aay358@my.utexas.edu,
	abhishek.nkulkarni21@gmail.com,
	utopcu@utexas.edu
}

\usepackage{acronym}
\usepackage{amsthm}
\usepackage{amsmath}
\usepackage{amssymb}
\usepackage{mathtools}
\usepackage{todonotes}
\usepackage{changes}
\usepackage{subcaption}
\usepackage{paralist}

\newcommand{\ie}{i.e.}

\newcommand{\lang}{\mathcal{L}}

\newcommand{\calA}{\mathcal{A}}

\newcommand{\truev}{\mathsf{true}}

\newcommand{\Eventually}{\Diamond \, }
\newcommand{\Next}{\bigcirc \, }
\newcommand{\until}{\mbox{$\, {\sf U}\,$}}

\newcommand{\weakpref}{\trianglerighteq}
\newcommand{\strictpref}{\triangleright}

\newcommand{\Paths}{\mathsf{Path}}

\newcommand{\swin}{\mathsf{SWin}}

\newcommand{\maximal}{\mathsf{Max}}
\newcommand{\minimal}{\mathsf{Min}}

\newtheorem{lemma}{Lemma}
\newtheorem{proposition}{Proposition}
\newtheorem{theorem}{Theorem}

\theoremstyle{definition}
 \newtheorem{assumption}{Assumption}
\newtheorem{definition}{Definition}
\newtheorem{problem}{Problem}

\newcommand{\refAlg}[1]{Algorithm~\ref{#1}}
\newcommand{\refAssume}[1]{Assumption~\ref{#1}}

\newcommand{\refDef}[1]{Definition~\ref{#1}}

\newcommand{\refFig}[1]{Figure~\ref{#1}}
\newcommand{\refLma}[1]{Lemma~\ref{#1}}

\newcommand{\refProp}[1]{Proposition~\ref{#1}}

\newcommand{\refThm}[1]{Theorem.~\ref{#1}}

\acrodef{mdp}[MDP]{Markov decision process}
\acrodef{pomdp}[POMDP]{Partially Observable Markov Decision Process}
\acrodef{momdp}[MOMDP]{Multi-objective MDP}
\acrodef{dfa}[DFA]{deterministic finite automaton}
\acrodef{tlmdp}[TLMDP]{terminating labeled Markov decision process}
\acrodef{lmdp}[LMDP]{labeled Markov decision process}
\acrodef{pdfa}[PDFA]{preference deterministic finite automaton}
\acrodef{pdra}[PDRA]{preference deterministic Rabin automaton}
\acrodef{cpltlf}[CPLTL$_f$]{Conditional Preference over LTL$_f$}
\acrodef{cpa}[CPA]{Conditional Preference Automaton}
\acrodef{scltl}[scLTL]{synctactically co-safe linear temporal logic}
\acrodef{ltl}[LTL]{linear temporal logic}
\acrodef{ltlf}[LTL$_f$]{linear temporal logic over finite traces}

\begin{document}
	
	\maketitle
	
	\begin{abstract}
		In multiplayer games with sequential decision-making, self-interested players form dynamic coalitions to achieve most-preferred temporal goals beyond their individual capabilities. We introduce a novel procedure to synthesize strategies that jointly determine which coalitions should form and the actions coalition members should choose to satisfy their preferences in a subclass of deterministic multiplayer games on graphs. In these games, a leader decides the coalition during each round and the players not in the coalition follow their admissible strategies. Our contributions are threefold. First, we extend the concept of admissibility to games on graphs with preferences and characterize it using maximal sure winning, a concept originally defined for adversarial two-player games with preferences. Second, we define a value function that assigns a vector to each state, identifying which player has a maximal sure winning strategy for certain subset of objectives. Finally, we present a polynomial-time algorithm to synthesize admissible strategies for all players based on this value function and prove their existence in all games within the chosen subclass. We illustrate the benefits of dynamic coalitions over fixed ones in a blocks-world domain. Interestingly, our experiment reveals that aligned preferences do not always encourage cooperation, while conflicting preferences do not always lead to adversarial behavior.
	\end{abstract}
	
	\section{Introduction}
    In multiplayer games involving sequential decision-making, self-interested players form dynamic coalitions to achieve objectives that exceed their individual capabilities.
Dynamic coalitions form to accomplish specific temporal sub-goals and then dissolve or reorganize into new coalitions to pursue subsequent sub-goals.
For instance, a team of waiter robots in a restaurant may coordinate briefly to rearrange a table before returning to their primary tasks of delivering orders. 
Investigating such multiplayer interactions with dynamic coalitions involves addressing two central questions \cite{osborne1994course}: How should a rational player decide which coalition to join? And what actions should members of a coalition collectively take to satisfy their individual temporal objectives?

Most existing research treats coalition formation and the synthesis of control strategies as separate problems.
Coalition games are the most widely studied models for coalition formation, which involve defining a preference structure over possible coalitions.
These preferences can be based on transferable utility (shared among members), non-transferable utility (individual and non-shareable), or qualitative rankings of coalitions by players (see \cite{Hajdukova2006} for a survey).

Rationality in coalition games is measured by stability—whether a player has an incentive to leave or join another coalition \cite{Hajdukova2006}.
While coalition games are traditionally studied in normal-form settings, recent work models them as dynamic Markovian processes, where players form coalitions sequentially and gain utility after each change \cite{Konishi-Ray}.
However, these models assume a predefined utility function and do not explicitly consider its dependence on the players' rational control strategies.

Literature on synthesizing control strategies has mainly focused on creating rational strategies for fixed coalitions of players. 
Rational synthesis \cite{Rational_synthesis} has focused on constructing strategies that satisfy temporal goals under solution concepts like Nash equilibrium or subgame perfect equilibrium, ensuring no player has an incentive to deviate from the formed coalition. 
On the other hand, approaches involving coalitional logic involves ATL* synthesis \cite{alur} and strategy logic \cite{StrategyLogic}, which model specific objectives for groups of players and synthesize strategies that guarantee these objectives are met collectively.
However, most studies assume that coalitions are predefined and stay constant during the interaction.

We address the problem of synthesizing a strategy that determines both the rational coalitions for the next round and the collective actions of their members in a deterministic multiplayer games on graphs.
Games on graphs \cite{gradel2003automata} are well-established models for sequential decision-making.  
In this paper, we assume that each player seeks to maximize the satisfaction of their partial-order preferences over a shared set of temporal logic objectives. 
As a result, the control strategy for each player is synthesized to maximally satisfy their preferences.

Preferences over temporal goals provide a rich,  realistic framework for decision-making in multi-agent systems \cite{bade2005nash,rahmani2024preference}. 
This framework enables us to examine how players can cooperate or compete to achieve their preferences. 
Decision-making with preferences over temporal goals poses two key challenges \cite{sen1997maximization}: \emph{Inescapability}, where players decide under limited information and time; and \emph{Incommensurability}, where some outcomes are incomparable. 
For example, decisions such as minimizing harm to one individual versus saving multiple lives, as in the trolley problem \cite{thomson1984trolley}, involve such trade-offs.

The game model we consider is based on the Stackelberg-Nash game \cite{Stackelberg-Nash}.
At the start of each round, a leader (player 1, $P_1$) decides and publicly announces the coalition for the next round, along with the actions for each member of that coalition. 
We assume that the actions of coalition members are rational, \ie, no member can achieve a better outcome by choosing a different action.
Players not in $P_1$'s coalition independently choose their actions based on their admissible strategies \cite{admissibility}.
For instance, in the waiter robots scenario, the restaurant manager, as the leader, assigns a group of robots to arrange tables, while the remaining robots continue delivering orders based on their individual strategies.

Synthesizing a strategy that jointly determines coalitions and actions for its members presents two main challenges.
First, which solution concept to use to analyze this game?
Given the interdependence between coalition formation and control strategy synthesis, existing solution concepts for coalition formation \cite{Konishi-Ray} and control strategy synthesis \cite{Rational_synthesis} are not suitable for the game model we consider.
In this model, to assess whether a coalition is rational, we must determine if each player can achieve a better outcome by joining the coalition instead of staying outside it.
On the other hand, a player's ability to secure a better outcome depends on forming a stable coalition that motivates other players to take actions that lead to this improvement.
Although recent work has explored synthesizing such joint strategies \cite{guelev}, it remains an underexplored problem in the literature.

Second, computing rational strategies in a multi-player game with partial-order preferences poses significant challenges.
Even for a single player, qualitative planning with partial-order preferences is known to be computationally hard \cite{rahmani2024preference,kulkarninash}. 
Moreover, they introduce additional complexity because they are combinative \cite{stanford}; an outcome in the game (\ie, a path in the game graph) can satisfy multiple goals simultaneously.
Planning with combinative preferences requires order lifting \cite{BARBERA1984185}---constructing a partial order on the powerset of temporal goals based on the partial order of individual temporal goals---which can be defined in multiple ways \cite{maly2020lifting}.
For instance, a cleaning robot may prioritize ``cleaning the living room'' over ``cleaning the bedroom,'' but if its battery allows, it can clean both rooms, achieving both goals. 
To the best of our knowledge, synthesis of rational strategies in multi-player game with partial-order preferences over temporal logic objectives have not been studied in the literature.

The key contributions of this paper are as follows.
\begin{itemize}
	\item \textbf{Solution concept.} 
	We extend the concept of admissibility to define the rationality of joint coalition formation and control strategies in deterministic multiplayer games on graphs with preferences.
	An admissible strategy ensures that no alternative strategy can lead to a strictly better outcome, no matter what strategies the other players choose.
	We prove that every maximal sure winning strategy is admissible.

	\item \textbf{Value function.} 
	We define a rank function and a value function that assign a vector to each state.
	The rank function, an order-preserving map \cite{davey2002introduction}, allows us to compare two states based on a player's preferences.
	The value function identifies which player has an admissible strategy to reach a state with a rank lower than a given value.
	In this way, the value function captures an undominated outcome each player can achieve independently of others' cooperation.
		
	\item \textbf{Synthesis algorithm.} 
	We introduce a novel algorithm that synthesizes an admissible strategy for $P_1$. 
	The admissible strategies for the other players can be constructed using the intermediate data structures generated by the algorithm. 
	We prove that an admissible strategy profile always exists for all players and that the algorithm constructs such a profile in polynomial time, based on the size of the product game.
 
\end{itemize}

Using a blocks-world domain \cite{nilsson2014principles}, we demonstrate that dynamic coalitions enable players to achieve better outcomes than fixed coalitions. 
Interestingly, our experiment also reveals a counter-intuitive scenario where aligned preferences do not always foster cooperation. By integrating coalition formation and preference satisfaction strategy synthesis, our work advances multi-player game theory, enabling algorithms that form coalitions based on players' preferences.
\
	
    \section{Preliminaries}
	\textbf{Notation.} 
Let $\Sigma$ be an alphabet (a finite set of symbols). We denote the set of finite (resp., infinite) words that can be generated using $\Sigma$ by $\Sigma^\ast$ (resp., $\Sigma^\omega$). 
Given a word $w\in \Sigma^\omega$, a prefix of $w$ is a word $u \in \Sigma^\ast$ such that there exists $v \in \Sigma^\omega$, $w=uv$.
A partition of a set $X$ is a set of subset $C_1 \ldots C_m$ of $X$ such that $C_1 \cup C_2 \cup \ldots \cup C_m = X$ and $C_i \cap C_j = \emptyset$ holds for all $i, j \in \{1, \ldots |X|\}$ and $i \neq j$.

Given a countable set $U$, a preference relation $\succeq$ on $U$ is \emph{preorder} on $U$. 
An element $u \in U$ is called \emph{maximal} if there is no $v \in U$ such that $u \succeq v$, and it is called \emph{minimal} if there is no $v \in U$ such that $v \succeq u$.
The sets of all maximal and minimal elements in $U$ are denoted by $\maximal(U, \succeq)$ and $\minimal(U, \succeq)$, respectively.

\subsection{Multiplayer Games on Graphs}

We consider a deterministic, multiplayer interaction modeled as a concurrent game on graph \cite{bouyer2011nash}.

\begin{definition}
	\label{def:multiplayer-game}
	 A deterministic multiplayer game on graph is a tuple,	
	$G=([N], S, \{A_{i}\}_{i \in [N]}, s_0, T, AP, L)$,
	where 
	\begin{inparaenum}[]
		\item $[N] = \{1,2,3,...,N\}$ is a finite, non-empty set of players.	
		\item $S$ is a finite, non-empty set of states.		
		\item  
        $A_i$ denotes the set of all actions available to the player $P_i$.  	
		\item $s_0 \in S$ is the initial state.
		\item $T: S \times A \rightarrow S$ is the deterministic transition function that maps a state and an action to the next state.	
		\item $AP$ is the set of atomic propositions. 	
		\item $L: S \rightarrow 2^{AP}$ is the labeling function that maps each state to a subset of atomic propositions.
	\end{inparaenum}
\end{definition}

A path in $G$ is an infinite sequence of states $\rho=s_0s_1s_2...$ where, for all $i \geq 0$, there exists $a \in A(s_i)$ such that $s_{k+1}=T(s_k, a)$. 
Given the labeling function $L$, every run $\rho$ in $G$ can be mapped to a word $w = L(\rho) = L(s_0)L(s_1)...$ over an alphabet $\Sigma = 2^{AP}$. 
The set of all paths in $G$ is denoted as $\mathsf{Paths}(G)$. 
A prefix of a path $\rho$ is a finite (non-empty) subsequence $\nu=s_0s_1s_2...s_k$ with length $|\nu|=k+1$. 
The set of all prefixes of a path $\rho$ is denoted as $\mathsf{Pref}(\rho)$ and the set of all prefixes in $G$ is $\mathsf{PrefPaths}(G)$. 

A finite-memory deterministic strategy $\pi_{i}: \mathsf{PrefPaths}(G) \rightarrow A_{i}$ of $P_i$ maps every non-empty prefix of $G$ to an action in $A_i$. 
The set of all strategies for $P_i$ is denoted as $\Pi_{i}$.
A strategy profile is a tuple of strategies, one for each player. 
Since the transition function and player strategies in $G$ are deterministic, every strategy profile $\pi = (\pi_1, \ldots, \pi_N)$ induces a unique path in $G$. We write $\mathsf{Path}_{G}(s_0, \pi)$ to denote the path defined by $\pi$.

Given a subset of players $C \subsetneq [N]$ and a strategy profile, we denote the set of joint actions available to players in $C$ by $A_C := \times_{i \in C} A_i$ and to those not in $C$ by $A_{-C} := \times_{i \notin C} A_i$.
Similary, we denote the joint strategies of players in $C$ by $\pi_C$ and that of players not in $C$ by $\pi_{-C}$.

\subsection{Specifying Temporal Goals}
\label{sec:ltlf}

We specify the temporal goals of players in $G$ using \ac{scltl} \cite{kupferman2001model}. 
Given a  set of atomic propositions $AP$, following grammar produces an \ac{scltl} formula:
\[
    \varphi \colon =  p \mid \neg p \mid \varphi  \land \varphi  \mid \Next \varphi \mid \varphi  \until \varphi \mid \varphi \lor \varphi,
\]

The temporal operators are used to specify properties of the system over sequences of time instants. The formula $\Next \varphi$ indicates that $\varphi$ holds true at the next time instant. Formula $\varphi_1 \until \varphi_2$ means there is future time instant at which $\varphi_2$ holds and at all instant from now until that instant, $\varphi_1$ holds true.
From those temporal operators, the temporal operator of $\Eventually$ (``Eventually'') is defined.
The formula $\Eventually \varphi$ means $\varphi$ holds true at a future time instant.
Formally, $\Eventually \varphi := \truev \until \varphi$. 

Every word satisfying an \ac{scltl} formula is characterized by a `good' prefix.
The set of good prefixes can be compactly represented as a language accepted by a finite automaton.

\begin{definition}
	A \ac{dfa} is a tuple $\calA = \langle Q, \Sigma, \delta, q_0, F \rangle$ where 
 	$Q$ is a finite state space.
 	$\Sigma$ is a finite alphabet. 
 	$\delta: Q \times \Sigma \rightarrow Q$ is a deterministic transition function.
    $q_0\in Q$ is an initial state, and $F\subseteq Q$ is a set of accepting (final) states.
\end{definition}

A transition from a state $q \in Q$ to a state $q' \in Q$ using input $\sigma \in \Sigma$ is denoted by $\delta(q, \sigma) = q'$. 
Slightly abusing the notation, we define the extended transition function $\delta: Q \times \Sigma^* \rightarrow Q$ as follows: $\delta(q, \sigma w) = \delta( \delta(q, \sigma), w )$ for each $w \in \Sigma^\ast$ and $\sigma \in \Sigma$, and $\delta(q, \epsilon) = q$ for each $q \in Q$, where $\epsilon$ is the empty string. The language of a \ac{dfa} $\calA$, denoted $\lang(\calA)$, consists of those words that induce a visit to an accepting state when input to the \ac{dfa}. Formally, $\lang(\calA) = \{ w \in \Sigma^* \mid \delta(q, w) \in F \}$. For every scLTL formula $\varphi$ over $AP$, there exists a \ac{dfa} such that $\lang(A_{\varphi}) = \lang(\varphi)$.

\subsection{Preference Modeling}

A preference model formally captures the notion of preferences in decision-making by representing the comparison between outcomes as a binary relation.
\begin{definition}
	\label{def:model_preference}
  	Given a countable set $U$, a preference model over $U$ is a tuple $\langle U, \weakpref \rangle$ in which $\weakpref$ is a \emph{preorder} on $U$, \ie, a reflexive and transitive binary relation on $U$.
\end{definition}

If the  preorder  $\weakpref$ on U  is also antisymmetric, it is said to be a \emph{partial order}. 
Let $\strictpref$ denote the irreflexive part of $\weakpref$. The set of maximal elements in $U$ under a preorder $\weakpref$ is the set $ \maximal(U, \weakpref) = \{u \in U \mid \nexists u' \in U: u' \weakpref u\}$. 
Similarly, the set of minimal elements in $U$ under a preorder $\weakpref$ is the set $\mathsf{Min}(U, \weakpref) = \{u \in U \mid \nexists u' \in U: u \weakpref u'\}$.
For every non-empty $U$, the sets $\maximal(U, \weakpref)$ and $\mathsf{Min}(U, \weakpref)$ are always non-empty \cite{sen1997maximization}. 
The elements of the set $\maximal(U, \weakpref)$ are called undominated elements of $U$.

A preference model $(\Phi, \weakpref)$ on a set of scLTL formulas induces a preference model on $(\Sigma^*, \weakpref)$ where for any $w_1, w_2 \in \Sigma^*$, $w_1 \weakpref w_2$ if and only if for every formula $\varphi \in \mathsf{MP}(w_1)$, there exists a formula $\varphi^{'} \in \mathsf{MP}(w_2)$ such that $\varphi \weakpref \varphi' $,
$w_1 \sim  w_2$ if and only if $\mathsf{MP}(w_1)=\mathsf{MP}(w_2)$ and
$w_1 \nparallel w_2$ otherwise, where $\mathsf{MP}(w):= \mathsf{Max}(\{\varphi \in \Phi \mid  w \models \varphi\}, \weakpref)$ denoting the undominated specifications that are satisfied when the path is $w$ \cite{rahmani2023probabilistic}.
	
	\section{Problem Formulation}
	
We consider deterministic multiplayer games on graphs where each player aims to maximally satisfy its own preference over a commonly shared set of scLTL formulas.

\begin{definition}
	\label{def:multiplayer-game-with-preferences}
		
	A deterministic multiplayer game on a graph with preferences is a tuple, $\langle G, \Phi, (\weakpref_{i})_{i \in [N]} \rangle,$ where $\Phi$ is a set of scLTL formulae and $\weakpref_{i}$ is the preference relation of player $P_i$'s on $\Phi$. 
\end{definition}

In multiplayer games on graphs with preferences (hereafter called a \emph{game}), coalition formation can allow players to improve their outcomes relative to their preferences when compared to playing individually \cite{hara2022coalitional}.

\begin{definition} 
	A \emph{coalition} $C$ is a non-empty subset of $[N]$. 
	A set of coalitions $\mathbb{C} = \{C_1, C_2, \ldots, C_m\}$ is called a \emph{coalition structure} if $\mathbb{C}$ is a partition of $[N]$.   
\end{definition}

A game is said to have \emph{dynamic coalitions} if the coalition structure changes during the game play. 
Analyzing these games involves investigating the following questions \cite{Konishi-Ray,Arnold-Schwarbe}:
\begin{enumerate}
	\item Who proposes the coalition structure in the next round?
	
	\item What constraints are imposed on coalition structures?
		
	\item How do players decide whether to accept the proposed coalition structure?	
\end{enumerate}

In this paper, we investigate a subclass of games inspired by Stackelberg Nash games \cite{Stackelberg-Nash}, where a leader determines the coalition structure and joint actions for the next round, and the remaining players follow the Nash equilibrium.
However, we adopt a different notion of rationality from the Nash equilibrium.

In our game model, player $P_1$ proposes a coalition structure and joint actions for the coalition members to execute in the next round.
Coalition members must accept the proposal if the proposed joint action is \emph{admissible} for them.
A coalition structure is considered admissible for a player if they have a strategy that guarantees an outcome no worse than opting out of any coalition \cite{brandenburger2008admissibility}.
In other words, $P_1$ determines the coalition structure for the next round, provided the proposal is admissible for all players.

We impose the following constraints on the coalition structures formed in the game. 

\begin{assumption}
	\label{assume:coalition-structure}
	
	Every coalition structure $\mathbb{C} = \{C_1, C_2, \ldots, C_m\}$ formed in a game satisfies one of the following conditions:
	\begin{inparaenum}[a)]
		\item Every coalition $C_j$ contains only one player, i.e., $|C_j| = 1$ for all $j \in [m]$.
		\item There exists a player $P_k$ such that $C_i \in  \{1, k\}$ for some $i \in [m]$, and for all other coalitions $C_j$ with $j \neq i$, we have $|C_j| = 1$.
	\end{inparaenum}
\end{assumption}

\refAssume{assume:coalition-structure} specifies that in every coalition structure formed in a game, $P_1$ either plays alone or forms a coalition with at most one other player, while all remaining players act independently.
This setup represents a situation where each player represents a unified group, eliminating the need to model larger coalitions.
For instance, in a university research lab, the advisor ($P_1$) might collaborate with individual student groups (each group is represented as a single player under \refAssume{assume:coalition-structure}) to expedite publishing multiple papers efficiently.

We now define some notation and formally define admissibility. Let $\mathbb{C}$ be the set of feasible coalitions in a game $G$ that satisfy \refAssume{assume:coalition-structure}. 
Let $\pi_1$ be a $P_1$ strategy that maps a state $s \in S$ to a coalition $C \in \mathbb{C}$ and a joint coalition action $a_C \in A_C$. 
The strategy for remaining players is a function $\pi_i$ for $i \in [N] \setminus \{1\}$ that maps a state $s \in S$, a coalition $C \in \mathbb{C}$, and a joint coalition action $a_C \in A_C$ to an action $a_i \in A_i$ of $P_i$. 
The set of all strategies of $P_i$ is denoted by $\Pi_i$ for $i \in [N]$.

Given two strategies $\pi_i, \pi_i' \in \Pi_i$ of player $P_i$, the strategy $\pi_i$ strictly dominates $\pi_i'$ if and only if, for every path $\rho \in \mathsf{Min}(\mathsf{Path}_G(s_0,\pi_i, \pi_{-i}))$, there exists a path $\rho' \in \mathsf{Min}(\mathsf{Path}_G(s_0,\pi_i', \pi_{-i}'))$ such that $L(\rho) \succ L(\rho')$,  where $\pi_{-i}, \pi_{-i}'$ are arbitrary joint strategies of all players except $P_i$.

Intuitively, a player strategy $\pi_i$ strictly dominates $\pi_i'$ if every least preferred outcome $\pi_i$ produces is strictly preferred to some least preferred outcome produced by $\pi_i'$, assuming the remaining players choose their available strategies arbitrarily.

\begin{definition}
	\label{def:admissible}
	A player $P_i$ strategy $\pi_i \in \Pi_i$ is admissible if it is not strictly dominated by any other strategy $\pi_i' \in \Pi_i$.
\end{definition}

We address the following problem in this paper.
\begin{problem} 
	\label{prob:problem}
	Given the game $\langle G,\Phi, (\weakpref_{i})_{i \in [N]} \rangle$, synthesize an admissible strategy profile in the game.
\end{problem}
	
	\section{Main Results}
	
We propose an algorithm to synthesize admissible strategies by leveraging maximal sure winning strategies \cite{kulkarni2025sequential}, a concept used to analyze two-player stochastic games on graphs with adversarial preferences. 
We prove that every maximal sure winning strategy is also an admissible strategy and use this insight to construct admissible strategies for all players.

\subsection{Maximal Sure Winning Strategy}

A maximal sure winning strategy ensures that a player achieves an outcome no worse than any outcome from any other strategy by that player \cite{kulkarni2025sequential}.
Synthesis of such strategies has been studied for two-player stochastic games on graphs, where players have partially ordered preferences over \ac{ltlf} objectives.
In contrast, we focus on deterministic multiplayer games on graphs.

\begin{definition}
	A player $P_i$ strategy $\pi_i$ is \emph{maximal sure winning} if, for every $P_i$ strategy $\pi_i'$ and every strategy profile $\pi_{-i}$ of the other players, the following condition holds: for every path $\rho \in \Paths(v_0, \pi_i, \pi_{-i})$, there exists a path $\rho' \in \Paths(v_0, \pi_i', \pi_{-i})$ such that $L(\rho) \succeq_i L(\rho')$.
\end{definition}

A maximal sure winning strategy for a player is an undominated strategy given the fixed strategies of the other players. 
In contrast, an admissible strategy does not assume fixed strategies for the other players.  
Maximal sure winning strategies are synthesized over a product game defined as follows.

\begin{definition}
	Given a game $\langle G, \Phi, (\weakpref_{i})_{i \in [N]} \rangle$ and preference automata corresponding to the relations $\weakpref_{i}$ for all $i \in [N]$ on $\Phi$, the product game is a tuple,
	$
		H= \langle V, \{A_{i}\}_{i \in [N]},\Delta, v_0, \mathbb{X}, (\mathcal{E}_{i})_{i \in [N]} \rangle,
	$
	where 
	\begin{inparaenum}[]
		\item $V = S \times Q$ is the set of states.
		\item $A_{i}$ is the set of actions of player $P_i$ as defined in \refDef{def:multiplayer-game}.
		\item $\Delta: V \times A \rightarrow V$ is a deterministic transition function.
		\item $v_0 = (s_0, \delta(q_0, L(s_0)))$ is an initial state, where $\delta$ is the transition function of preference semi-automaton.
		\item $\mathcal{E}_{i}$ is a partial order on a partition $\mathbb{X}$ of $V$, where every node $X \in \mathbb{X}$ represents a subset of states in $V$. 
	\end{inparaenum}
\end{definition}

The product game encodes the preference relation over a set of \ac{scltl} formulas as a preference relation over partitions of states in the product game.
This encoding allows us to monitor progress toward satisfying all formulas in $\Phi$, enabling the comparison of two paths in the product game according to each player's preference relation.

Maximal sure winning strategies can be synthesized in polynomial-time using an order-preserving map called \emph{rank}.

\begin{definition}
	\label{def:rank}
	
	Given a preorder $\succeq$ on a set $U$, let $Z_0 = \maximal(U, \succeq)$ and, for all $k \geq 0$, let
	\[
	Z_{k+1} = \maximal\left(U \setminus \bigcup_{j=0}^{k} Z_j, \succeq \right).
	\]
	The rank of any state $u \in U$, denoted by $\text{rank}_{\succeq}(u) = k$, is the smallest integer $k \geq 0$ such that $u \in Z_k$.
\end{definition}

\refDef{def:rank} assigns a unique, finite rank to every state in $A$ given a preorder $\succeq$. 
Since the set $\maximal(W, \succeq)$ is non-empty for any non-empty subset $W \subseteq U$, the inductive assignment of ranks terminates only when the subset $U \setminus \bigcup_{j=0}^{k} Z_j$ is empty, i.e., when a rank has been assigned to all states in $U$. 
Additionally, the sets $Z_0, Z_1, \ldots$ are mutually exclusive and exhaustive subsets of $U$. 
Therefore, every state in $U$ has a unique rank under a given preorder.

Every partial order can admit multiple order-preserving maps \cite{davey2002introduction}. 
\refDef{def:rank} employs an order-preserving map that enables the synthesis of maximal sure winning strategies.

In fact, rank is a partial map that possesses the following properties \cite{kulkarni2025sequential}. 
First, if two states, $v$ and $v'$, have the same rank then they are either equally preferred or incomparable. 
Second, if $v$ has a higher rank than $v'$, then $v$ cannot be less preferred than $v'$. 
Finally, if $v$ is strictly more preferred than $v'$, then $v$ must have a lower rank than $v'$.
The converse statements of these properties are not always true. 

A maximal sure winning strategy can be understood in terms of rank. 
The rank of a strategy for $P_i$ is determined by the minimal outcome it guarantees, regardless of the strategies chosen by the other players.
Formally, the rank of a strategy is defined as:
\begin{align*}
	\mathsf{MaxRank}_i(&\pi_i) = \\ &\max\{\mathsf{Rank}_{i}(\mathsf{Path}_{G}(s_0,\pi_i,\pi_{-i}))|\pi_{-i} \in \Pi_{-i}\}.
\end{align*}
A strategy $\pi_i$ is a maximal sure winning strategy if no alternative strategy $\pi_i'$ satisfies $\mathsf{MaxRank}_i(\pi_i') < \mathsf{MaxRank}_i(\pi_i)$.

We now present the main result of this section that highlights the connection between maximal sure winning and admissibility. 
First, we prove a lemma.

\begin{lemma}
    \label{lemma1:min}
	Given a player $P_i$ strategy $\pi_i$, $\mathsf{MaxRank}_i(\pi_i)=k$ holds only if there exists a path ${\rho \in \mathsf{Min}\{\mathsf{Paths}(\pi_i)\}}$ such that $\mathsf{Rank}_{i}(w) = k$.
\end{lemma}

The proofs of \refLma{lemma1:min} and upcoming results in this paper are available in the supplementary material.

\begin{proposition}
\label{prop1-sure-admissible}
	Every maximal sure winning strategy of $P_i$ in $H$ is admissible for $P_i$, for any $i \in [N]$.
\end{proposition}
\begin{proof}
	By contradiction. 
	Suppose that there exists a $P_i$ strategy $\pi_i'$ that dominates its maximal sure winning strategy $\pi_i$.
	Then, by definition, for every minimal path $\rho'$ in $\mathsf{Min}\{\mathsf{Paths}(\pi_i')\}$, there exists a minimal path $\rho \in \mathsf{Min}\{\mathsf{Paths}(\pi_i)\}$ such that ${\rho' \succ_i \rho}$.
	But \refLma{lemma1:min} implies that a path that attains a rank ${\mathsf{MaxRank}_i(\pi_i)}$ is a minimal path in the set ${\mathsf{Min}\{\mathsf{Paths}(\pi_i)\}}$. 
	Therefore, we have $\mathsf{MaxRank}_i(\pi_i) > \mathsf{MaxRank}_i(\pi_i')$ which means that $\pi_i$ is not a maximal sure winning strategy, a contradiction. 
\end{proof}

\subsection{Value of a State}

The rank of a maximal sure winning strategy for a player is the smallest ranked outcome that the player can achieve without cooperation.
We define the \emph{value of a state} to be the vector of ranks of maximal sure winning strategies of all players,
\begin{align*}
    \mathsf{Val}(v) = [\mathsf{MaxRank}_i(\pi_i) \mid i \in [N]],
\end{align*}
where $\pi_i$ is maximal sure winning strategy of player $P_i$. 

\refAlg{alg:max-swin} presents a procedure to compute the value of each state in $H$. 
The algorithm adapts \cite[Alg. 1]{kulkarni2025sequential}, which is defined for two-player stochastic game with adversarial preferences, to a multiplayer deterministic game with preferences. 
The following theorem states that this adaptation is sound.

\begin{algorithm}[t]
	\caption{Value assignment for player-$i$}\label{alg:max-swin}
	\begin{algorithmic}[1]
		\STATE \textbf{Input:} Product game $H$ and $P_i$.
		\STATE Initialize $\mathsf{Val}(v) \gets \infty$ for all $v \in V$. 
		\FORALL{$k = 0 \ldots R_i^{\max}$}	
			\STATE 	$Y_k \gets \{v \in V \mid \mathsf{Rank}_i(v) \leq k\}$
			\STATE $U_k \gets \mathsf{SureWin}_i(Y_k)$

			\STATE Set $\mathsf{Val}(v) \gets \min \{k, \mathsf{Val}(v)\}$ for all $v \in U_k$.

		\ENDFOR
		\RETURN $\mathsf{Val}$
	\end{algorithmic}
\end{algorithm}

\begin{theorem} 
    \label{Thm1-Alg1}

	The value $\mathsf{Val}_i(v)$ computed by \refAlg{alg:max-swin} for any state $v \in V$ is the smallest rank that $P_i$ can guarantee achieving, assuming the remaining players are adversarial. 
\end{theorem}

\subsection{Synthesis of Admissible Strategies} 

\refAlg{alg:synthesis} outlines a procedure to synthesize an admissible strategy for $P_1$. It specifies what coalitions to form at each state and the corresponding actions to ensure an undominated outcome for $P_1$. 
The intermediate data structures generated by the algorithm facilitate the synthesis of admissible strategies for the remaining players.

\begin{algorithm}[t]
	\caption{Admissible Actions under Dynamic Coalitions}
	\label{alg:synthesis}
	\begin{algorithmic}[1]
		\REQUIRE Product game, $H$, value function $\mathsf{Val}$, rank $l$.
		\STATE $V_{0} \gets \{v \in V \mid 
		\mathsf{Rank}_{1}(v) \leq l\}$
            \STATE $\mathsf{CoalV}(v) \gets \mathsf{Val}(v)$ for all $v \in V$. 
		\REPEAT 
            \STATE $k \gets 0$, $U_k \gets \mathsf{Pre}(V_k) \setminus V_k$
		\FOR{$v \in U_k$}
		\STATE Initialize $\mathsf{CandV}$ be a map from a state and a coalition action to admissible actions and their values for non-coalition members. 
		\FOR{$a_{C} \in \mathsf{Act}(V_k, v)$}
            \STATE Let $C = \{1, i\}$ if $|C| > 1$.
		\IF{$|C| > 1 \land \mathsf{Val}_i(v) <  \max\{\mathsf{CoalV}_i(\Delta(v,a_C))\}$}
		\STATE $\mathsf{Act}(V_k, v) \gets \mathsf{Act}(V_k, v) \setminus \{a_C\}$ 
		\STATE \textbf{continue}
		\ELSIF{$|C| > 1$}
		\STATE $c_i \gets \max\{\mathsf{CoalV}_i(\Delta(v, a_C))\} $
		\ENDIF
		\FOR {$j \in [N]\setminus C$}
		\STATE $\mathrel{c_j \gets \min\limits_{a_j\in A_j(v)} \max\{\mathsf{CoalV}_i(\Delta(v,a_C,a_j)\}}$
		\STATE $a_j \gets \arg \min\limits_{a_j\in A_j(v)} \max\{\mathsf{CoalV}_i(\Delta(v,a_C,a_j)\}$
		\ENDFOR
            \STATE $a_{-C} \gets (a_j)_{j \in [N]\setminus C}$
            \STATE $\mathsf{CandV}[v, a_C, a_{-C}] \gets  (c_j)_{j \in [N]\setminus C}$
 
		\ENDFOR
		\IF{$ \mathsf{Act}(V_k, v)=\emptyset$}
		\STATE $ U_{k} \gets U_{k} \setminus \{v\}$ 
		\ELSE
		\FOR{$\mathrel{k \in [N] \setminus \{1\}}$}
		\STATE $C_k \gets \max (\{c_k \mid \forall a_C, \forall a_{-C}:  c \in \mathsf{CandV}[v, a_C, a_{-C}]\})$
		\ENDFOR
		\STATE $\mathsf{CoalV}(v) \gets (C_2, C_3,..., C_N)$ 
		\ENDIF
		\ENDFOR
		\STATE $V_{k+1} \gets V_k \cup U_{k}$
		\STATE $k \gets k + 1$
		\UNTIL {$V_{k+1} = V_{k}$}
		\IF{$v_0 \notin V_k$}
		\RETURN No admissible strategy exists.
		\ENDIF
		\RETURN $\mathsf{Act}(V_k), \mathsf{CandV}$
	\end{algorithmic}
\end{algorithm}

Given a rank $l$, \refAlg{alg:synthesis} determines if $P_1$ has a strategy to reach a state with rank no greater than $l$ from the initial state of $H$. 
If such a strategy exists, the algorithm returns the set of admissible actions for each state and a map, $\mathsf{CandV}$.
The dictionary $\mathsf{CandV}$ maps every state, potential coalition, and joint actions (coalition and non-coalition) to the values each player would achieve if that action \((a_C, a_{-C})\) is executed in state \( v \).
To find an undominated strategy for $P_1$, we repeatedly invoke \refAlg{alg:synthesis} with \( l \) going from $0$ to \( r^{1}_{\text{max}} \).

The \texttt{for} loop between lines \texttt{5} and \texttt{24} constructs a level set of states. 
From any state at level \( k+1 \) of this level set, $P_1$ can enforce a visit to a state at level \( k \), either independently or by forming a coalition. 
Specifically, given the subset of states up to level \( V_k \), the loop examines all states in \( U_k \) where a joint action \( (a_C, a_{-C}) \) can induce a visit to \( V_k \). 

The \texttt{if}-block between lines \texttt{9-14} implicitly (through the omitted else case) examines each state in \( U_k \) to determine all actions that allow $P_1$ to independently enforce a visit to \( V_k \). 
These actions are deemed admissible for $P_1$.
It also evaluates every possible coalition from \( v \).
Line \texttt{9} checks whether a coalition’s joint action is rational by verifying if the transition increases the value of the state for the coalition member $P_i$. 
If the value increases, the coalition and its action are deemed inadmissible and removed from the set \( \mathsf{Act}(V_k, v) \), which maps each state in \( U_k \) to the set of valid actions from \( v \) to reach \( V_k \). 
Otherwise, the value $P_i$ receives from the transition is recorded in the \texttt{CandV} map. 
If \( \mathsf{Act}(V_k, v) \) is empty, the algorithm removes \( v \) from \( U_k \). 
Given a potential coalition $C = \{1, i\}$ and their joint action $(a_1, a_i)$ the minimum value that each player can individually guarantee from state $v$ given $C$ and $(a_1, a_i)$ is also recorded in \texttt{CandV} map.

An admissible strategy for $P_1$ selects any action from $\mathsf{Act}(V_k, v)$ at state $v$. 
For $P_i$ (\(i \in [N] \setminus \{1\}\)), an admissible strategy selects a value-minimizing action using \texttt{CandV} map given a state and the joint action of the coalition. 
The vector $\mathsf{CoalV(v)}$ represents the worst-case values of a state $v$, with $\mathsf{CoalV_i(v)}$ for $P_i$, based on all possible choices of rational coalition actions from $\mathsf{Act}(V_k, v)$ by $P_1$ (Line \texttt {2}).

Next, we prove the correctness of \refAlg{alg:synthesis}.
First, we define what constitutes a rational action for a coalition $\{1, i\}$.

\begin{definition} 
	\label{def:rationality} 
	Given a state $v \in H$ and a coalition $\{1, i\}$, the action $(a_1,a_i)$ is rational for $P_i$ if, for any choice of actions of remaining player $a_{-1-i}$, the value $\mathsf{CoalV}_i(v')$ of the resulting state $v' = \Delta(v, a_1, a_i, a_{-1-i})$ is not strictly greater than $\mathsf{Val}_i(v)$.
\end{definition}

The following proposition establishes the consistency of value updates.

\begin{proposition}
\label{prop2:induction}
	In $k$-th iteration of \refAlg{alg:synthesis}, for any state $v \in U_k$, player $P_i$ has an admissible strategy to visit a state with rank smaller than or equal to $\mathsf{CoalV}_i(v)$. 
\end{proposition}

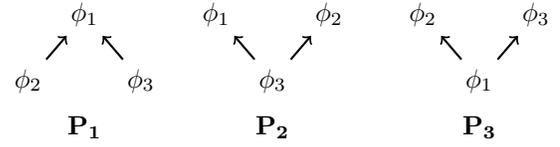
\begin{figure}[tb]
    \centering
    \begin{tikzpicture}[scale=0.5]
        
        \node (s1) at (-6.5, 1.5) {$\phi_1$};
        \node (s2) at (-8, -0.25) {$\phi_2$};
        \node (s3) at (-5, -0.25) {$\phi_3$};

        \node at (-6.5, -1.5) {$\mathbf{P_1}$};

        \draw[->, thick] (s2) -- (s1); 
        \draw[->, thick] (s3) -- (s1);

        \node (s4) at (-3, 1.5) {$\phi_1$};
        \node (s5) at (0, 1.5) {$\phi_2$};
        \node (s6) at (-1.5, -0.25) {$\phi_3$};

        \draw[->, thick] (s6) -- (s4); 
        \draw[->, thick] (s6) -- (s5); 

         \node at (-1.5, -1.5) {$\mathbf{P_2}$};

        \node (s7) at (5.5, 1.5) {$\phi_3$};
        \node (s8) at (2.5, 1.5) {$\phi_2$};
        \node (s9) at (4, -0.25) {$\phi_1$};

        \node at (4, -1.5) {$\mathbf{P_3}$};

        \draw[->, thick] (s9) -- (s7); 
        \draw[->, thick] (s9) -- (s8);

    \end{tikzpicture}
    \caption{Preferences over temporal objectives for arm P1, P2, and P3, where $\phi_1 = \Diamond (e_1 \land \Diamond e_2), \phi_2 = \neg e_2 \mathsf{U} e_3,$ and $\phi_3 = \Diamond (e_3 \land \Diamond e_2)$.}
    \label{fig:prefs}
\end{figure}

The following lemma establishes that \refAlg{alg:synthesis} only discards actions that are not rationalizable.

\begin{lemma} 
\label{lemma2:rationalizable}
	At any iteration \( k \), \refAlg{alg:synthesis} discards a joint action if and only if it is not rationalizable.
\end{lemma}

Finally, we establish that the strategy constructed above is indeed admissible for $P_1$.

\begin{theorem} 
    \label{Thm2:admissible}
	The dynamic coalition strategy \( \pi \) synthesized based on \refAlg{alg:synthesis} is an admissible strategy for $P_1$.
\end{theorem}
\begin{proof}
Suppose that \refAlg{alg:synthesis} terminates at rank \( l \leq r^i_{\max} \). 
	By \refLma{lemma2:rationalizable}, for any \( l' < l \), no sequence of rationalizable joint actions can enforce a transition from \( v_0 \) to a set of states with rank at most \( l' \) for \( P_1 \). Therefore, any other strategy \( \pi' \) must satisfy \( \mathsf{MaxRank}_1(\pi) \leq \mathsf{MaxRank}_1(\pi') \), implying that \( \pi \) is a maximal sure-winning strategy. 
	By \refProp{prop1-sure-admissible}, \( \pi \) is admissible.  
\end{proof}

\textbf{Complexity.} The time complexity of Algorithm 2 depends on the number of iterations and the actions it evaluates at each step. The algorithm iterates at most \( |V| r^1_{\max} \) times, where \( |V| \) is the total number of states (excluding the initial state \( v_0 \)) and \( r^1_{\max} \) is the maximum rank for $P_1$. For each state \( v \), the algorithm considers all possible actions of $P_1$, (\( |A_1| \)), and all joint actions of $P_1$ and any other player $P_j$, (\( \sum_{j=2}^N |A_j||A_1| \)). This results in a worst-case time complexity of \( \mathcal{O}(r^1_{\max} |V| ( |A_1| + \sum_{j=2}^N |A_j||A_1|)) \).

	\section{Experiments}
	\begin{figure*}
    \centering
    \begin{tikzpicture}[scale=0.5]

\node[anchor=north west, font= ] at (-0.5, 6) {S0};
\draw[thick] (-0.5,-0.5) rectangle (4.5,6); 
\node[anchor=north east, font=\tiny ] at (4.5, 6) {Label: $\emptyset$};
\node[anchor=north east, font=\tiny ] at (4.6, 5.5) {$\mathsf{Val}:[3,3,3]$};
\node[anchor=north east, font=\tiny ] at (4.6, 5) {$\mathsf{CoalV}:[0,1]$};
\draw[thick] (1.5, 4) rectangle (4.5,6);

\draw[fill=brown!40] (0,-0.25) rectangle (3,0.5); 
\draw (1, -0.25) -- (1, 0.5);
\node at (0.5, 0.12) {1};
\draw (2, -0.25) -- (2, 0.5);
\node at (1.5, 0.12) {2};
\node at (2.5, 0.12) {3};

\draw[fill=blue!30] (0,0.5) rectangle (1,1.5); 
\node at (0.5, 1) {B1};

\draw[fill=red!30] (0,1.5) rectangle (1,2.5); 
\node at (0.5, 2) {B2};

\draw[fill=yellow!30] (0,2.5) rectangle (1,3.5); 
\node at (0.5, 3) {B3};

\draw[fill=yellow!30] (0,3.5) rectangle (1,4.5); 
\node at (0.5, 4) {B4};

\draw[->, thick] (4.6, 2.75) -- (6.5, 2.75);

\node at (5.5,3.25) {$\{1,2\}$};

\node[anchor=north west, font= ] at (6.5, 6) {S1};
\draw[thick] (6.5,-0.5) rectangle (12.5,6);

\draw[fill=brown!40] (7,-0.25) rectangle (10,0.5); 
\draw (8, -0.25) -- (8, 0.5);
\node at (7.5, 0.12) {1};
\draw (9, -0.25) -- (9, 0.5);
\node at (8.5, 0.12) {2};
\node at (9.5, 0.12) {3};

\draw[fill=yellow!30] (7,0.5) rectangle (8,1.5); 
\node at (7.5, 1) {B3};

\draw[fill=yellow!30] (7,1.5) rectangle (8, 2.5); 
\node at (7.5, 2) {B4};

\draw[thick] (9.9,3) -- (10.9,3); 
\draw[thick] (9.3,3.6) arc[start angle=90, end angle=-90, radius=0.6]; 
\node at (11.4, 3) {P1};

\draw[fill=blue!30] (8.5,2.5) rectangle (9.5,3.5); 
\node at (9, 3) {B1};

\draw[thick] (9.9,1.5) -- (10.9,1.5); 
\draw[thick] (9.3,2.1) arc[start angle=90, end angle=-90, radius=0.6]; 
\node at (11.4, 1.5) {P2};

\draw[fill=red!30] (8.5,1) rectangle (9.5,2); 
\node at (9, 1.5) {B2};

\node[anchor=north east, font=\tiny] at (12.5, 5.5) {$\mathsf{Val}: [3,3,3]$};
\node[anchor=north east, font=\tiny] at (12.5, 5) {$\mathsf{CoalV}: [0.1]$};
\node[anchor=north east, font=\tiny] at (12.5, 6) {Label: $\emptyset$};

\draw[thick] (9.5, 4) rectangle (12.5,6);

\draw[->, thick] (12.6, 2.75) -- (14.5, 2.75);

\node at (13.5, 3.25) {$\{1,2\}$};

\node[anchor=north east, font=\tiny] at (19.5, 5.5) {$\mathsf{Val}: [3,3,3]$};
\node[anchor=north east, font=\tiny] at (19.5, 5) {$\mathsf{CoalV}: [0.1]$};
\node[anchor=north east, font=\tiny] at (19.5, 6) {Label:$\{e_1\}$};
\node[anchor=north west, font= ] at (14.5, 6) {S2};

\draw[thick] (14.5,-0.5) rectangle (19.5,6);

\draw[thick] (16.5, 4) rectangle (19.5,6); 
\draw[fill=brown!40] (15,-0.25) rectangle (18,0.5); 

\draw (16, -0.25) -- (16, 0.5);
\node at (15.5, 0.12) {1};
\draw (17, -0.25) -- (17, 0.5);
\node at (16.5, 0.12) {2};
\node at (17.5, 0.12) {3};

\draw[fill=yellow!30] (15,0.5) rectangle (16,1.5); 
\node at (15.5, 1) {B3};

\draw[fill=yellow!30] (15,1.5) rectangle (16,2.5); 
\node at (15.5, 2) {B4};

\draw[fill=blue!30] (16,0.5) rectangle (17,1.5); 
\node at (16.5, 1) {B1};

\draw[fill=red!30] (17,0.5) rectangle (18,1.5); 
\node at (17.5, 1) {B2};

\draw[->, thick] (19.6, 2.75) -- (21.5, 2.75);

\node at (20.5, 3.25) {$\{1,3\}$};

\node[anchor=north east, font=\tiny] at (28.5, 5.5) {$\mathsf{Val}: [3,3,3]$};
\node[anchor=north east, font=\tiny] at (28.5, 5) {$\mathsf{CoalV}: [0.1]$};
\node[anchor=north east, font=\tiny] at (28.5, 6) {Label: $\emptyset$};

\node[anchor=north west, font= ] at (21.5, 6) {S3};

\draw[thick] (21.5,-0.5) rectangle (28.5,6);

\draw[fill=brown!40] (22,-0.25) rectangle (25,0.5); 
\draw (23, -0.25) -- (23, 0.5);
\node at (22.5, 0.12) {1};
\draw (24, -0.25) -- (24, 0.5);
\node at (23.5, 0.12) {2};
\node at (24.5, 0.12) {3};

\draw[fill=yellow!30] (22,0.5) rectangle (23,1.5); 
\node at (22.5, 1) {B3};

\draw[fill=yellow!30] (22,1.5) rectangle (23, 2.5); 
\node at (22.5, 2) {B4};

\draw[thick] (25.4, 4) rectangle (28.5,6);

\draw[thick] (24.4,3.6) arc[start angle=90, end angle=-90, radius=0.6]; 
\node at (26.5, 3) {P1};
\draw[thick] (25,3) -- (26,3);
\draw[fill=blue!30] (23.5,2.5) rectangle (24.5,3.5); 
\node at (24, 3) {B1};

\draw[thick] (24.4,2.1) arc[start angle=90, end angle=-90, radius=0.6]; 
\node at (26.5, 1.5) {P2};
\draw[thick] (25,1.5) -- (26,1.5);
\draw[fill=red!30] (23.5,1) rectangle (24.5,2); 
\node at (24, 1.5) {B2};

\draw[->, thick] (28.6, 2.75) -- (30.4, 2.75);

\node at (29.5, 3.25) {$\{1,2\}$};

\node[anchor=north west, font= ] at (30.4, 6) {S4};

\draw[thick] (30.4,-0.5) rectangle (34.9,6); 
\node[anchor=north east, font=\tiny] at (34.9, 6) {Label: $\{e_2\}$};
\node[anchor=north east, font=\tiny] at (34.9, 5.5) {$\mathsf{Val}: [0,0,1]$};
\node[anchor=north east, font=\tiny] at (34.9, 5) {$\mathsf{CoalV}= [0,1]$};

\draw[thick] (31.5, 4) rectangle (34.9,6);

\draw[fill=brown!40] (30.8,-0.25) rectangle (33.8,0.5); 
\draw (31.8, -0.25) -- (31.8, 0.5);
\node at (31.3, 0.12) {1};
\draw (32.8, -0.25) -- (32.8, 0.5);
\node at (32.3, 0.12) {2};
\node at (33.3, 0.12) {3};

\draw[fill=yellow!30] (30.8,0.5) rectangle (31.8,1.5); 
\node at (31.3, 1) {B3};

\draw[fill=yellow!30] (30.8,1.5) rectangle (31.8,2.5); 
\node at (31.3, 2) {B4};

\draw[fill=blue!30] (32.8,0.5) rectangle (33.8,1.5); 
\node at (33.3, 1) {B1};

\draw[fill=red!30] (32.8,1.5) rectangle (33.8,2.5); 
\node at (33.3, 2) {B2};

\end{tikzpicture}
    \caption{The first dynamic coalition formation strategy that achieves the most preferred outcome for $P_1$}
    \label{fig:first-path}
\end{figure*}
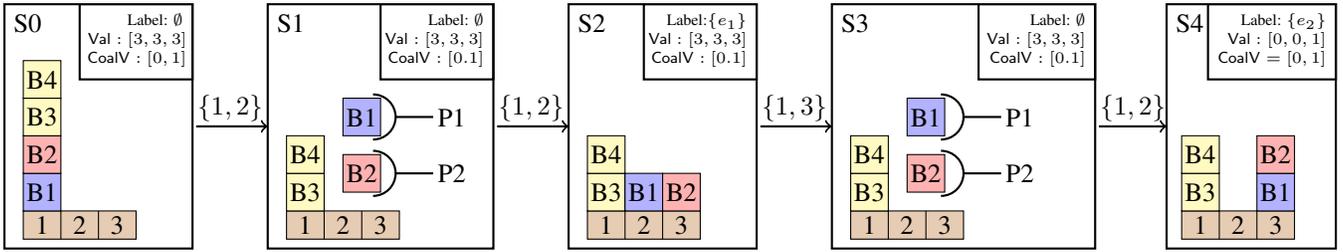

\begin{figure*}[ht!]
\centering
\begin{tikzpicture}[scale=0.5]

\node[anchor=north west, font= ] at (-0.5, 6) {S0};
\draw[thick] (-0.5,-0.5) rectangle (4.5,6);

\node[anchor=north east, font=\tiny] at (4.5, 6) {Label:$\emptyset$};
\node[anchor=north east, font=\tiny] at (4.6, 5.5) {$\mathsf{Val}:[3,3,3]$};
\node[anchor=north east, font=\tiny] at (4.6, 5) {$\mathsf{CoalV}:[0,1]$};

\draw[thick] (1.5, 4) rectangle (4.5,6);

\draw[fill=brown!40] (0,-0.25) rectangle (3,0.5);
\draw (1, -0.25) -- (1, 0.5);
\node at (0.5, 0.12) {1};
\draw (2, -0.25) -- (2, 0.5);
\node at (1.5, 0.12) {2};
\node at (2.5, 0.12) {3};
\draw (2, 0) -- (2, 0.5);

\draw[fill=blue!30] (0,0.5) rectangle (1,1.5); 
\node at (0.5, 1) {B1};

\draw[fill=red!30] (0,1.5) rectangle (1,2.5); 
\node at (0.5, 2) {B2};

\draw[fill=yellow!30] (0,2.5) rectangle (1,3.5); 
\node at (0.5, 3) {B3};

\draw[fill=yellow!30] (0,3.5) rectangle (1,4.5); 
\node at (0.5, 4) {B4};

\draw[->, thick] (4.6, 2.75) -- (6.5, 2.75);

\node at (5.5, 3.25) {$\{1,3\}$};

\node[anchor=north west, font= ] at (6.5, 6) {S1}; 
\draw[thick] (6.5,-0.5) rectangle (12.5,6); 

\draw[fill=brown!40] (7,-0.25) rectangle (10,0.5); 
\node at (7.5, 0.12) {1};
\draw (8, -0.25) -- (8, 0.5);
\node at (8.5, 0.12) {2};
\node at (9.5, 0.12) {3};
\draw (9, -0.25) -- (9, 0.5);

\draw[fill=red!30] (7,0.5) rectangle (8,1.5); 
\node at (7.5, 1) {B2};

\draw[fill=yellow!30] (7,1.5) rectangle (8, 2.5); 
\node at (7.5, 2) {B3};

\draw[thick] (10,3) -- (11,3);

\draw[thick] (9.4,3.6) arc[start angle=90, end angle=-90, radius=0.6]; 
\node at (11.5, 3) {P1};

\draw[fill=blue!30] (8.5,2.5) rectangle (9.5,3.5); 
\node at (9, 3) {B1};

\draw[thick] (10,1.5) -- (11,1.5);

\draw[thick] (9.4,2.1) arc[start angle=90, end angle=-90, radius=0.6]; 
\node at (11.5, 1.5) {P3};

\draw[fill=yellow!30] (8.5,1) rectangle (9.5,2); 
\node at (9, 1.5) {B4};

\node[anchor=north east, font=\tiny] at (12.6, 5.5) {$\mathsf{Val}: [3,3,3]$};

\node[anchor=north east, font=\tiny] at (12.5, 6) {Label:$\emptyset$};
\node[anchor=north east, font=\tiny] at (12.6, 5) {$\mathsf{CoalV}:[0,0]$};

\draw[thick] (9.5, 4) rectangle (12.5,6);

\draw[->, thick] (12.6, 2.75) -- (14.5, 2.75);

\node at (13.5, 3.25) {$\{1,3\}$};

\node[anchor=north west, font= ] at (14.5, 6) {S2}; 
\draw[thick] (14.5,-0.5) rectangle (19.5,6);

\node[anchor=north east, font=\tiny] at (19.6, 6) {Label:$\{e_1\}$};
\node[anchor=north east, font=\tiny] at (19.6, 5.5) {$\mathsf{Val}: [3,3,3]$};
\node[anchor=north east, font=\tiny] at (19.6, 5) {$\mathsf{CoalV}:[0,0]$};

\draw[thick] (16.5, 4) rectangle (19.5,6);

\draw[fill=brown!40] (15,-0.25) rectangle (18,0.5); 
\draw (16, -0.25) -- (16, 0.5);
\node at (15.5, 0.12) {1};
\draw (17, -0.25) -- (17, 0.5);
\node at (16.5, 0.12) {2};
\node at (17.5, 0.12) {3};
\draw (18, -0.25) -- (18, 0.5);

\draw[fill=red!30] (15,0.5) rectangle (16,1.5); 
\node at (15.5, 1) {B2};

\draw[fill=yellow!30] (15,1.5) rectangle (16,2.5); 
\node at (15.5, 2) {B3};

\draw[fill=blue!30] (15,2.5) rectangle (16,3.5); 
\node at (15.5, 3) {B1};

\draw[fill=yellow!30] (17,0.5) rectangle (18,1.5); 
\node at (17.5, 1) {B4};

\draw[->, thick] (19.6, 2.75) -- (21.5, 2.75);

\node at (20.5, 3.25) {$\{1,2\}$};

\node[anchor=north west, font= ] at (21.5, 6) {S3}; 
\draw[thick] (21.5,-0.5) rectangle (28.5,6); 

\draw[fill=brown!40] (22,-0.25) rectangle (25,0.5); 
\draw (23, -0.25) -- (23, 0.5);
\node at (22.5, 0.12) {1};
\draw (24, -0.25) -- (24, 0.5);
\node at (23.5, 0.12) {2};
\node at (24.5, 0.12) {3};

\draw[fill=red!30] (22,0.5) rectangle (23,1.5); 
\node at (22.5, 1) {B2};

\draw[fill=yellow!30] (22,1.5) rectangle (23, 2.5); 
\node at (22.5, 2) {B3};

\draw[thick] (25,3) -- (26,3);

\draw[thick] (24.4,3.6) arc[start angle=90, end angle=-90, radius=0.6]; 
\node at (26.5, 3) {P1};

\draw[fill=blue!30] (23.5,2.5) rectangle (24.5,3.5); 
\node at (24, 3) {B1};

\draw[thick] (25,1.5) -- (26,1.5);

\draw[thick] (24.4,2.1) arc[start angle=90, end angle=-90, radius=0.6]; 
\node at (26.5, 1.5) {P3};

\draw[fill=yellow!30] (23.5,1) rectangle (24.5,2); 
\node at (24, 1.5) {B4};

\node[anchor=north east, font=\tiny] at (28.6, 5.5) {$\mathsf{Val: [3,3,3]}$};

\node[anchor=north east, font=\tiny] at (28.5, 6) {Label:$\emptyset$};
\node[anchor=north east, font=\tiny] at (28.6, 5) {$\mathsf{CoalV}:[0,0]$};

\draw[thick] (25.5, 4) rectangle (28.5,6);

\draw[->, thick] (28.6, 2.75) -- (30.5, 2.75);
\node at (29.5, 3.25) {$\{1,3\}$};

\node[anchor=north west, font= ] at (30.4, 6) {S4}; 
\draw[thick] (30.5,-0.5) rectangle (35,6);

\node[anchor=north east, font=\tiny] at (35, 6) {Label:$\{e_1,e_3\}$};
\node[anchor=north east, font=\tiny] at (35, 5.5) {$\mathsf{Val}: [0,0,0]$};
\node[anchor=north east, font=\tiny] at (35, 5) {$\mathsf{CoalV}: [0,0]$};

\draw[thick] (31.5, 4) rectangle (35,6);

\draw[fill=brown!40] (31,-0.25) rectangle (34,0.5);
\draw (32, -0.25) -- (32, 0.5);
\node at (31.5, 0.12) {1};
\draw (33, -0.25) -- (33, 0.5);
\node at (32.5, 0.12) {2};
\node at (33.5, 0.12) {3};

\draw[fill=red!30] (31,0.5) rectangle (32,1.5); 
\node at (31.5, 1) {B2};

\draw[fill=yellow!30] (31,1.5) rectangle (32,2.5);
\node at (31.5, 2) {B3};

\draw[fill=blue!30] (33,0.5) rectangle (34,1.5); 
\node at (33.5, 1) {B1};

\draw[fill=yellow!30] (33,1.5) rectangle (34,2.5); 
\node at (33.5, 2) {B4};

\end{tikzpicture}
\caption{The second dynamic coalition formation strategy that achieves the most preferred outcome for $P_1$}
\label{fig:second-path}
\end{figure*}
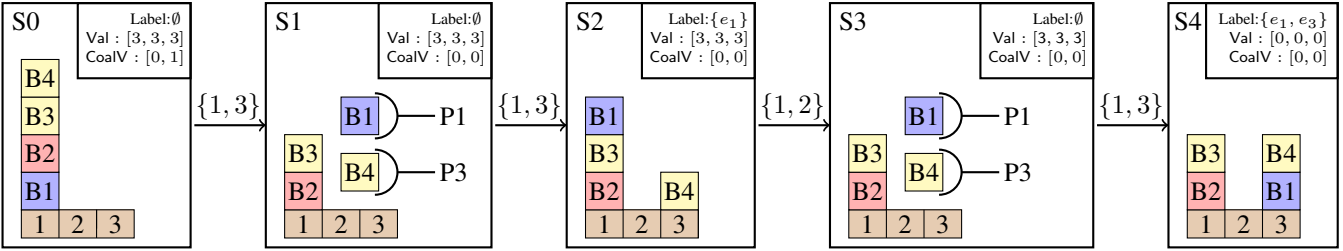

We demonstrate the advantage of dynamic coalitions over static coalitions and present a counter-intuitive coalition formation strategy with an experiment in an adapted version of the BlocksWorld domain \cite{nilsson2014principles}.

Our experiment involves three robot arms ($P_1$, $P_2$, $P_3$) controlling four blocks (B1, B2, B3, B4) to achieve temporal objectives. At any time, blocks are either on the table (brown in Figure 2) or held by an arm. The table has three stackable locations. $P_1$, as the leader, controls only B1, $P_2$ controls B2, and $P_3$ controls B3 and B4. An arm may choose to do nothing, pick up a block assigned to it, or place a held block at one of the locations. If multiple arms try to place blocks at the same location, $P_1$ has priority, followed by $P_2$, then $P_3$. Once $P_1$ declares a coalition and a joint action, non-members act according to their admissible strategies, with priority considered where necessary.

We define three state labels:
\begin{inparaenum}[]
\item $e_1$: (B1 is on B3) or (B1 and B2 are at locations 2 and 3).
\item $e_2$: Two blocks are stacked at location 3.
\item $e_3$: B4 is on top of B1.
\end{inparaenum}

The preferences of each arm over shared scLTL formulas are shown in \refFig{fig:prefs} 

\begin{align*}
\Phi = \{\Diamond (e_1 \land \Diamond e_2), \neg e_2 \mathsf{U} e_3, \Diamond (e_3 \land \Diamond e_2) \}
\end{align*}

\subsection{Advantages of Dynamic Coalitions}

In static coalitions, $P_1$ is a member of a fixed coalition throughout the game, while in dynamic coalitions, $P_1$ can adjust the coalition each round. 
We hypothesized that dynamic coalitions would guarantee a better outcome for $P_1$, \ie, a smaller rank in the worst-case, compared to static coalitions. 
Figures \ref{fig:first-path} and \ref{fig:second-path} show two paths that result when players follow their admissible strategies from the initial state S0.

Figure \ref{fig:first-path} shows that a static coalition between $P_1$ and $P_2$ can only enforce a rank $3$ outcome for $P_1$ by visiting S2 (see $\mathsf{Val}_1(\text{S2})$).
After this state, $P_1$ must prevent $P_3$ from picking B3 or B4 to avoid stacking two blocks at location 3. 
Thus, $P_1$ has to switch to a coalition with $P_3$. 
This coalition is rational for $P_3$ as the value of S2 for $P_3$ is 3, as it cannot enforce any specification by itself from S2, whereas the further coalitional strategy of $P_1$ guarantees $\Diamond (e_1 \land \Diamond e_2)$ to be satisfied, which is of rank 1 for $P_3$, (see $\mathsf{CoalV}(S3)$).  

Similarly, a fixed coalition between $P_1$ and $P_3$ cannot enforce the specification $\Diamond (e_1 \land \Diamond e_2)$ without cooperation from $P_2$. As shown in Figure \ref{fig:second-path}, while $P_1$ and $P_3$ can reach S2 where label $e_1$ occurs, they must place B1 and B4 on location 3 to satisfy $\Diamond (e_1 \land \Diamond e_2)$  and $\Diamond (e_3 \land \Diamond e_2)$. However, $P_2$'s action can disrupt this, so $P_1$ must form a coalition with $P_2$ to control the outcome and guarantee $\Diamond (e_1 \land \Diamond e_2)$ to be satisfied in the next round.

These examples show that dynamic coalition is necessary for P1 to achieve its most preferred outcome.

\subsection{Preference Alignment and Coalition Formation}

Intuitively, we expect players with aligned preferences to form more coalitions while those with conflicting preferences to form coalitions less often \cite{van2009coalition}. 
Our experiment demonstrates that this intuition is not valid.

Figure \ref{fig:first-path} illustrates a path that supports our intuition. However, Figure \ref{fig:second-path} shows an instance where $P_1$ can also achieve $\Diamond (e_1 \land \Diamond e_2)$  by forming more coalitions with $P_3$ even though $P_1$ and $P_3$ have conflicting preferences. While intermediate cooperation with $P_2$ is still needed, this behavior suggests that the level of alignment between preferences do not directly relate to frequency of a coalition.

Interestingly, the second path also optimizes the outcome for all agents, satisfying all specifications. While our analysis does not explicitly prioritize social welfare in synthesized strategies, this example shows how coalition strategies can improve overall outcomes.

	\section{Conclusion}
	We consider the sequential decision making problem on deterministic multiplayer games on graphs where self-interested players aim to maximally satisfy temporal goals by forming temporary coalitions. In contrast to previous work on coalition formation games, we combine coalition formation strategy and control strategy into a single strategy for the leader of the game. We employed rank as an order preserving map to represent the worst-case performance of a strategy under incomplete preferences as well as the rationalizability of a coalition. We introduced an algorithm for synthesizing an admissible strategy for the leader that guarantees to achieve minimum rank in the worst case. We demonstrated in a BlocksWorld domain experiment that using the dynamic coalition formation strategy we synthesized enables the leader to achieve a better outcome compared to sticking with the same coalition throughout the entire game play. We also showed that our algorithm has polynomial-time complexity.
Building on results of this paper, we consider a number of future directions including  assessing the relationship between admissible coalition formation strategies and social welfare and generalizing our approach to stochastic games. 
	
	\appendix

	\section*{Acknowledgments}
    This work was supported in part by ONR N00014-22-1-2703, ARL W911NF-23-2-0011.

    \bibliographystyle{named}
    \bibliography{aaai25}
    
    \appendix
    \section{Omitted Proofs}
    \begin{proof}[Proof (\refLma{lemma1:min})]
    By contradiction. 
	Suppose that ${\mathsf{MaxRank}_i(\pi_i)=k}$ and 
	there exists a path ${\rho \in \mathsf{Paths}(\pi_i)}$ such that $\mathsf{Rank}_i(\rho)=k$ and $\rho \notin \mathsf{Min}\{\mathsf{Paths}(\pi_i)\}$. 
	Since $\rho$ is not a minimal path, there exists a minimal path $\rho'$ that satisfies $\rho \succ_i \rho'$.
	By definition, we have $k = \mathsf{Rank}_{i}(\rho) < \mathsf{Rank}_{i}(\rho') \leq \mathsf{MaxRank}_i(\pi_i)$, a contradiction.
\end{proof}

\begin{proof}[Proof (\refThm{Thm1-Alg1})]
The value of a state $v \in V$ is updated on line 6, if and only if $v$ is a sure winning state for player $i$ to reach $Y_k$. 
This update applies to all states in $V$ because, by the property of sure-winning computation, $v \in \swin_i(Y)$ whenever $v \in Y$. 
It is possible that a state $v$ is included in $U_k$ for multiple values of $k \geq 0$. 
Particularly, if $v \in U_k$ for some $k \geq 0$, then $v \in U_K$ for all $K \geq k$, because $\swin_i(Y) \subseteq \swin_i(Y')$ holds whenever $Y \subseteq Y'$. 
In this case, the minima operator ensures that the value assigned to $v$ is the smallest $k$ for which $v$ is winning, preserving the minimality of the computed solution.
\end{proof}

\begin{proof} [Proof (\refProp{prop2:induction})]

By induction on $k$. 
	
	\textbf{(Base case).} 
	At $k=0$,  $\mathsf{CoalV}_i(u)$ is assigned $\mathsf{Val}_i(u)$ (Line \texttt{2}). 
	By definition, $P_i$ can enforce a visit to a state with rank at most \( \mathsf{Val}_i (u)\) by following its maximal sure-winning strategy.
	Since every maximal sure-winning strategy of $P_i$ is also admissible, $P_i$ has an admissible strategy to reach a state with rank at most \( \mathsf{CoalV}_i(u)\).  
	
	\textbf{(Inductive step).}
	Assume the proposition holds for \( k = 1, \ldots, m \). For $u \in U_{m+1}$, the value of \( \mathsf{CoalV}_{i}(u) \) for $P_i$ is updated based on two cases.
	
	(Case 1). \emph{When $P_i$ is in coalition with $P_1$:} 
	In this case, \( \mathsf{CoalV}_{i}(u) \) is the maximum of \( \mathsf{CoalV}_{i}(u') \), where \( u' \) is a state in \( U_m \) reachable from \( u \) under some joint action \( (a_1, \ldots, a_N) \).
	By the induction hypothesis, $P_i$ is guaranteed to achieve a rank less than or equal to \( \mathsf{CoalV}_{i}(u') \) from the subsequent state \( u' \) by following an admissible strategy. Therefore, $P_i$ is ensured to achieve a rank less than or equal to \( \mathsf{CoalV}_{i}(u) \) by following its admissible strategy.
	
	(Case 2). \emph{When $P_i$ is not in coalition with $P_1$:}
	In this case, \( \mathsf{CoalV}_{i}(u) \) is the minimum of the maximum values it can attain over \( \mathsf{CoalV}_{i}(u') \), where \( u' \) is a state in \( U_m \) reachable from \( u \). 
	This occurs when $P_i$ fixes its action \( a_i \) (Line \texttt{16}), the coalition \( C \) selects \( a_C \), and the remaining players choose arbitrary actions. 
	By the induction hypothesis, $P_i$ is guaranteed to achieve a rank less than or equal to \( \mathsf{CoalV}_{i}(u') \) from \( u' \) by following an admissible strategy. 
	Therefore, $P_i$ is ensured to achieve a rank less than or equal to \( \mathsf{CoalV}_{i}(u) \) by following its admissible strategy.
\end{proof}

\begin{proof} (\refLma{lemma2:rationalizable})

\textbf{($\implies$)} Let \( C = \{1, i\} \). 
	If \refAlg{alg:synthesis} discards a joint action \( a_C \), then the value of $P_i$ at state \( v \), denoted \( \mathsf{Val}_i(v) \), is less than the minimum possible value under dynamic coalitions after taking action \( a_C \), i.e., \( \mathsf{Val}_i(v) < \max \{\mathsf{CoalV}_i(\Delta(v, a_C))\} \). 
	By \refProp{prop2:induction}, this minimum value corresponds to the worst-case scenario for $P_i$. 
	Thus, after taking action \( a_C \), $P_i$ cannot enforce a transition to a set of states with rank at most \( \mathsf{Val}_i(v) \), making \( a_C \) not rationalizable.
	
	\textbf{($\impliedby$)} Suppose there exists a minimum iteration \( k \) where a joint action \( a_C \in \mathsf{Act}(V_k, v) \) is not discarded, but is not rationalizable. 
	By the minimality of \( k \), all previously considered joint actions are rationalizable. 
	\refProp{prop2:induction} ensures that the worst-case scenario for $P_i$ is exactly \( \max \{\mathsf{CoalV}_i(\Delta(v, a_C))\} \). 
	If \( a_C \) is not rationalizable, then \( \mathsf{Val}_i(v) < \max \{\mathsf{CoalV}_i(\Delta(v, a_C))\} \), meaning \refAlg{alg:synthesis} must discard \( a_C \) at line \texttt{10}. 
	This contradicts the assumption that \( a_C \) was not discarded. 
	
	Thus, \refAlg{alg:synthesis} discards a joint action if and only if it is not rationalizable.  
\end{proof}

\end{document}